\documentclass[copyright,creativecommons]{eptcs}

\usepackage{breakurl}             
\usepackage{graphicx}
\usepackage{alltt}

\title{Developing Experimental Models for NASA Missions with ASSL}
\author{
Emil Vassev
\institute{Lero -- The Irish Software Engineering Research Centre\\University College Dublin, Ireland}
\email{Emil.Vassev@lero.ie}
\and
Mike Hinchey
\institute{Lero -- The Irish Software Engineering Research Centre\\University of Limerick, Ireland}
\email{Mike.Hinchey@lero.ie}
}

\begin{document}
\maketitle

\begin{abstract}
NASA's new age of space exploration augurs great promise for deep space exploration missions whereby spacecraft should be independent, autonomous, and smart. Nowadays NASA increasingly relies on the concepts of autonomic computing, exploiting these to increase the survivability of remote missions, particularly when human tending is not feasible. Autonomic computing has been recognized as a promising approach to the development of self-managing spacecraft systems that employ onboard intelligence and rely less on control links. The Autonomic System Specification Language (ASSL) is a framework for formally specifying and generating autonomic systems. As part of long-term research targeted at the development of models for space exploration missions that rely on principles of autonomic computing, we have employed ASSL to develop formal models and generate functional prototypes for NASA missions. This helps to validate features and perform experiments through simulation. Here, we discuss our work on developing such missions with ASSL.  
\end{abstract}

\section{Introduction}
Autonomic Computing (AC) is an emerging field for the development of large-scale self-managing complex systems \cite{ac-book}. The idea behind is that complex systems such as spacecraft can autonomously manage themselves and deal with dynamic requirements and unanticipated threads. NASA is currently approaching AC with interest, recognizing in its concepts a promising approach to developing new class of space exploration missions, where spacecraft should be independent, autonomous, and smart. Here, AC software makes spacecraft autonomic systems capable of planning and executing many activities onboard the spacecraft to meet the requirements of changing objectives and harsh external conditions. Examples of such AC-based unmanned missions are the Autonomous Nano-Technology Swarm (ANTS) concept mission \cite{nasa-swrm-msns} and the Deep Space One mission \cite{ac-book}. It is important to mention though that the development of systems implying such a large-scale automation implies sophisticated software, which requires new development approaches and new verification techniques. 

Practice has shown that traditional development methods cannot guarantee software reliability and prevent software failures, which is very important in complex systems such as spacecraft where software errors can be very expensive and even catastrophic (e.g., the malfunction in the control software of Ariane-5 \cite{model-check}). Thus, formal methods need to be employed in the development of autonomic spacecraft systems. When employed correctly, formal methods have proven to be an important technique that ensures quality of software \cite{ten-cmmndmnts}. In the course of this research, to develop models for autonomic space exploration missions, we employ the ASSL (Autonomic System Specification Language) formal method \cite{assl-book},  \cite{assl-computer}. Conceptually, ASSL is an AC-dedicated framework providing a powerful formal notation and computational tools that help developers with problem formation, system design, system analysis and evaluation, and system implementation.

\section{Preliminaries}
\subsection{Targeted NASA Missions}
Both NASA ANTS and NASA Voyager missions are targeted by this research. Other space-exploration missions, such as NASA Mars Rover and ESA's Herschel are a subject of interest as well. 

\subsubsection{NASA ANTS}
The ANTS (Autonomous Nano-Technology Swarm concept sub-mission PAM (Prospecting Asteroids Mission) is a novel approach to asteroid-belt resource exploration. ANTS provides extremely high autonomy, minimal communication requirements to Earth, and a set of very small explorers with a few consumables \cite{nasa-swrm-msns}. The explorers forming the swarm are pico-class, low-power, and low-weight spacecraft, yet capable of operating as fully autonomous and adaptable agents. 
There are three classes of ANTS spacecraft: \textit{rulers}, \textit{messengers} and \textit{workers}. By grouping them in certain ways, ANTS forms teams that explore particular asteroids. The internal organization of a team depends on the task to be performed and on the current environmental conditions. In general, each team has a group leader (ruler), one or more messengers, and a number of workers carrying a specialized instrument. The messengers are needed to connect the team members when they cannot connect directly.

\subsubsection{NASA Voyager}
\begin{figure}[htp]
	\begin{centering}
	\includegraphics[width=0.6\textwidth]{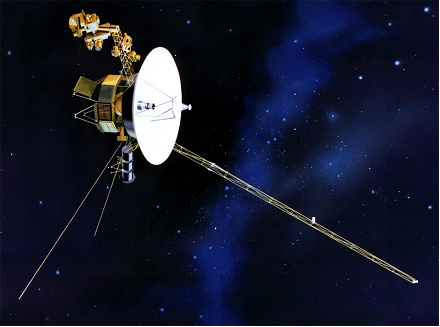}
	\caption{ Voyager Spacecraft \cite{nasa-voyager}}
	\label{fig:voyager}
  \end{centering}
\end{figure}

The NASA Voyager Mission \cite{nasa-voyager} was designed for exploration of the Solar System. The mission started in 1977, when the twin spacecraft Voyager I and Voyager II were launched (cf. Figure \ref{fig:voyager}). The original mission objectives were to explore the outer planets of the Solar System. As the Voyagers flew across the Solar System, they took pictures of planets and their satellites and performed close-up studies of Jupiter, Saturn, Uranus, and Neptune.

After successfully accomplishing their initial mission, both Voyagers are now on an extended mission, dubbed the "Voyager Interstellar Mission". This mission is an attempt to chart the heliopause boundary, where the solar winds and solar magnetic fields meet the so-called \textit{interstellar medium} \cite{nasa-voyager-intrstllr}.

\subsection{ASSL}
The ASSL framework \cite{assl-book}, \cite{assl-computer} provides a powerful formal notation and suitable mature tool support that allow ASSL specifications to be edited and validated and Java code to be generated from any valid specification. ASSL is based on a specification model exposed over hierarchically organized formalization tiers. This specification model is intended to provide both infrastructure elements and mechanisms needed by an autonomic system (AS). The latter is considered as being composed of special \textit{autonomic elements} (AEs) interacting over \textit{interaction protocols}, whose specification is distributed among the ASSL tiers. Note that each tier is intended to describe different aspects of the AS in question, such as \textit{service-level objectives}, \textit{policies}, \textit{interaction protocols}, \textit{events}, \textit{actions}, etc. This helps to specify an AS at different levels of abstraction imposed by the ASSL tiers.

The following elements represent the major tiers and sub tiers in ASSL. 

\begin{flushleft}
I. Autonomic System (AS)
\end{flushleft}
\begin{itemize}
\setlength{\itemsep}{0pt}%
\setlength{\parskip}{0pt}%
	\item AS Service-level Objectives
	\item AS Self-managing Policies
	\item AS Architecture
	\item AS Actions
	\item AS Events
	\item AS Metrics
\end{itemize}
\begin{flushleft}
II. AS Interaction Protocol (ASIP)
\end{flushleft}
\begin{itemize}
\setlength{\itemsep}{0pt}%
\setlength{\parskip}{0pt}%
	\item AS Messages
	\item AS Communication Channels
	\item AS Communication Functions
\end{itemize}
\begin{flushleft}
III. Autonomic Element (AE)
\end{flushleft}
\begin{itemize}
\setlength{\itemsep}{0pt}%
\setlength{\parskip}{0pt}%
	\item AE Service-level Objectives
	\item AE Self-managing Policies
	\item AE Friends
	\item AE Interaction Protocol (AEIP)
	\begin{itemize}
	\setlength{\itemsep}{0pt}%
  \setlength{\parskip}{0pt}%
		\item AE Messages
		\item AE Communication Channels
		\item AE Communication Functions
		\item AE Managed Elements
	\end{itemize}
	\item AE Recovery Protocol
	\item AE Behavior Models
	\item AE Outcomes
	\item AE Actions
	\item AE Events
	\item AE Metrics
\end{itemize}

As shown, the ASSL multi-tier specification model decomposes an AS in two directions:
\begin{enumerate}
\item into levels of functional abstraction;
\item into functionally related tiers (sub-tiers).
\end{enumerate}

With the first decomposition, an AS is presented from three different perspectives, these depicted as three main tiers (main concepts): 
\begin{itemize}
\item AS Tier forms a general and global AS perspective exposing the architecture topology, general system behaviour rules, and global \textit{actions}, \textit{events} and \textit{metrics} applied to these rules.
\item ASIP Tier (AS interaction protocol) forms a communication perspective exposing a means of communication for the AS under consideration. 
\item AE Tier forms a unit-level perspective, where an interacting sets of the AS's individual components is specified. These components are specified as AEs with their own behaviour, which must be synchronized with the behaviour rules from the global AS perspective.
\end{itemize}

\section{Research}
In this section, we present our research objectives and current trends.

\subsection{Objectives}
This research emphasizes the ASSL formal development approach to autonomic systems (ASs). We believe that ASSL may be successfully applied to the development of experimental models for space-exploration missions integrating autonomic features. Thus, we use ASSL to develop experimental models for NASA missions in a stepwise manner (feature by feature) and generate a series of prototypes, which we evaluate in simulated conditions. Here, it is our understanding that both prototyping and formal modeling, which will aid in the design and implementation of real space-exploration missions, are becoming increasingly necessary and important as the urgent need emerges for higher levels of assurance regarding correctness.

\subsection{Benefits for Space Systems}
Experimental modeling of space-exploration missions can be extremely useful for the design and implementation of such systems. The ability to compare features and issues with actual missions and with hypothesized possible autonomic approaches gives significant benefit. In our approach, we develop space-mission models in a series of incremental and iterative steps where each model includes new autonomic features. This helps to evaluate the performance of each feature and gradually construct a model of more realistic space exploration missions. Different prototypes can be tried and tested (and benchmarked as well), and get valuable feedback before we implement the real system. Moreover, this approach helps to discover eventual design flaws in existing missions and the prototype models. 

\subsection{Modeling NASA ANTS and NASA Voyager with ASSL}
ASSL has been successfully used to specify autonomic features and generate prototype models for two NASA projects - the ANTS (Autonomous Nano-Technology Swarm) concept mission (cf. Section 2.1.1) and the Voyager mission (cf. Section 2.1.2). In both cases the generated prototype models helped to simulate space-exploration missions and validate features through simulated experimental results. 

In our endeavor to develop NASA missions with ASSL, we emphasized modeling ANTS self-managing policies \cite{twrds-assl-ants} of self-configuring, self-healing and self-scheduling and the Voyager image-processing autonomic behavior \cite{assl-voyager}. In addition, we proposed a specification model for the ANTS safety requirements. In general, a complete specification of these autonomic properties requires a two-level approach. They need to be specified at the individual spacecraft level (AE tier) and at the level of the entire system (AS tier). Here, to specify the self-managing policies we used four base ASSL elements:
\begin{itemize}
\item	\textit{a self-managing policy structure} - which describes the self-managing policy under consideration. We use a set of special ASSL constructs such as \textit{fluents} and \textit{mappings} to specify such a policy \cite{assl-book}. With fluents we express specific situations, in which the policy is interested, and with mappings we map those situations to actions. 
\item	\textit{actions} - a set of actions that can be undertaken by ANTS in response to certain conditions, and according to that policy.
\item	\textit{events} - a set of events that initiate fluents and are prompted by the actions according to the policies.
\item	\textit{metrics} - a set of metrics \cite{assl-book} needed by the events and actions.
\end{itemize}

\begin{figure}[t]
\begin{minipage}[t]{0.480\textwidth}
\begin{alltt} \scriptsize
\hrulefill
\textbf{SELF_PROTECTING} \{
 \textbf{FLUENT} inSecurityCheck \{
  \textbf{INITIATED_BY} \{ \textbf{EVENTS}.privateMessageIsComming \}
  \textbf{TERMINATED_BY} \{ \textbf{EVENTS}.privateMessageSecure, 
                  \textbf{EVENTS}.privateMessageInsecure \}\}
 \textbf{MAPPING} \{
  \textbf{CONDITIONS} \{ inSecurityCheck\}
  \textbf{DO_ACTIONS} \{ \textbf{ACTIONS}.checkPrivateMessage \}\}\}
\hrulefill
\end{alltt}
\vspace{-6mm}
  \caption{Self-managing Policy}
\vspace{-2mm}
  \label{fig:assl-policy-example}
\end{minipage}
\hfill
\begin{minipage}[t]{0.500\textwidth}
\begin{alltt} \scriptsize
\hrulefill
\textbf{EVENT} privateMessageIsComming \{
 \textbf{ACTIVATION} \{ \textbf{SENT} \{ \textbf{AEIP.MESSAGES}.privateMessage \}\}\} 
\textbf{EVENT} privateMessageInsecure \{
 \textbf{GUARDS} \{ \textbf{NOT METRICS}.thereIsInsecureMsg \}
 \textbf{ACTIVATION} \{ \textbf{CHANGED} \{ \textbf{METRICS}.thereIsInsecureMsg \}\}\} 
\textbf{EVENT} privateMessageSecure \{
 \textbf{GUARDS} \{ \textbf{METRICS}.thereIsInsecureMsg \}
 \textbf{ACTIVATION} \{ \textbf{CHANGED} \{ \textbf{METRICS}.thereIsInsecureMsg \}\}\} 
\hrulefill
\end{alltt}
\vspace{-6mm}
\caption{Policy Events}
\vspace{-2mm}
\label{fig:policy_events}
\end{minipage}
\begin{minipage}[t]{0.900\textwidth}
\begin{alltt} \scriptsize
\hrulefill
\textbf{ACTION} checkPrivateMessage \{ \textbf{GUARDS} \{ .... \} \textbf{ENSURES} \{ .... \}
 \textbf{DOES} \{	senderIdentified = \textbf{call ACTIONS}.checkSenderCertificate; ....	\}
 \textbf{ONERR_DOES} \{ .... \}	\textbf{TRIGGERS} \{ .... \} \textbf{ONERR_TRIGGERS} \{ .... \} \} 
\hrulefill
\end{alltt}
\vspace{-6mm}
\caption{Action}
\vspace{-3mm}
\label{fig:policy_action}
\end{minipage}
\end{figure}

Figure \ref{fig:assl-policy-example}, Figure \ref{fig:policy_events},
and Figure \ref{fig:policy_action} present a partial specification of a \textit{self-protecting policy} employed by one of the prototype models we built for the NASA ANTS concept mission. Note that ASSL events (cf. Figure \ref{fig:policy_events}) and actions (cf. Figure \ref{fig:policy_action}) may be specified with a special \textit{GUARDS} clause stating preconditions that must be met before an event may be raised or an action may be undertaken. In addition, events (cf. Figure \ref{fig:policy_events}) are specified with a special \textit{ACTIVATION} clause and actions may be specified with an \textit{ENSURES} clause to state post-conditions that must be met after the action execution. Actions may call other actions in their \textit{DOES} or \textit{ONERR\_DOES} clauses. Finally, actions (cf. Figure \ref{fig:policy_action}) may trigger events specified in special \textit{TRIGGERS} and \textit{ONERR\_TRIGGERS} clauses. Note that the \textit{ONERR\_DOES} and \textit{ONERR\_TRIGGERS} clauses specify the action execution path in case of an error \cite{assl-book}.

\subsection{Formal Verification}
Safety is a major concern to NASA missions, where both reliability and maintainability form an essential part. In that context, the ASSL framework toolset provides verification mechanisms for automatic reasoning about a specified AS. The base validation approach in ASSL comes in the form of consistency checking. The latter is a mechanism for verifying ASSL specifications by performing exhaustive traversal to check for both syntax and consistency errors (type consistency, ambiguous definitions, etc.). In addition this mechanism checks whether a specification conforms to special correctness properties, defined as ASSL semantic definitions. Although considered efficient, the ASSL consistency checking mechanism cannot handle logical errors (specification flaws) and thus, it is not able to assert safety (e.g., freedom from deadlock) or liveness properties. 

Currently, a model-checking validation mechanism able to handle such errors is under development \cite{assl-modelcheck}. The ASSL model-checking mechanism is the next generation of the ASSL consistency checker based on automated reasoning. By allowing for automated system analysis and evaluation, this mechanism completes the AS development process with ASSL. The ability to verify the ASSL specifications for design flaws can lead to significant improvements in both specification models and generated ASs. Subsequently, ASSL can be used to specify, validate and generate better prototype models for current and future space-exploration systems. 

In general, model checking provides an automated method for verifying finite state systems by relying on efficient graph-search algorithms. The latter help to determine whether or not system behavior described with temporal correctness properties holds for the system's state graph. In ASSL, the model-checking problem is: given autonomic system $A$ and its ASSL specification $a$, determine in the system's state graph $g$ whether or not the behavior of $A$, expressed with the correctness properties $p$, meets the specification $a$. Formally, this can be presented as a triple $(a,p,g)$ where: $a$ is the ASSL specification of the autonomic system $A$, $p$ presents the correctness properties specified in FOLTL, and $g$ is the state graph constructed from the ASSL specification in a labeled transition system (LTS) \cite{model-check} format.

However, due to the so-called state-explosion problem \cite{model-check}, model checking cannot efficiently handle logical errors in large ASSL specifications. Therefore, to improve the detection of errors introduced not only with the ASSL specifications, but also with the supplementary coding, the automatic verification provided by the ASSL tools is augmented by appropriate testing. Currently, a novel test generator tool based on change-impact analysis is under development. This tool helps the ASSL framework automatically generate high-quality test suites for self-managing policies. The test generator tool accepts as input an ASSL specification comprising sets of policies that need to be tested and generates a set of tests each testing a single execution path of a policy.

\section{Conclusion}
In this research, we place emphasis on modeling autonomic properties of space-exploration missions with ASSL. With ASSL we model such properties through the specification of self-managing policies and service-level objectives. Formal verification handles consistency flaws and operational Java code is generated for valid models. Generated code is the basis for functional prototypes of space-exploration missions employing self-managing features. Those prototypes may be extremely useful when undertaking further investigation based on practical results and help developers test the autonomic behavior under simulated conditions.

\bibliographystyle{eptcs} 
\bibliography{developing_nasa_models_assl}

\end{document}